\documentclass[11pt,a4paper]{article}

\RequirePackage{ifpdf} 
\usepackage{amsmath} 
\usepackage{mathtools}

\usepackage{jheppub}
\usepackage{pstricks}
\usepackage[final]{pdfpages} 
\usepackage{ifpdf} 
\usepackage{slashed}
\usepackage{hyperref}

\allowdisplaybreaks

\usepackage{color} 
\usepackage{graphics}

\usepackage{etoolbox} 
\usepackage{fixmath}

\usepackage{tikz}
\usepackage[compat=1.1.0]{tikz-feynman}
\usetikzlibrary{decorations.pathmorphing}
\usetikzlibrary{positioning,arrows}
\usetikzlibrary{decorations.markings}
\usetikzlibrary{shapes,shadows}
\usepackage{float}

\usepackage{amsfonts}

\usepackage{epstopdf}

\usepackage[normalem]{ulem}
\usepackage{framed}
\usepackage{comment}
\usepackage{longtable}
\usepackage{comment}
\usepackage{booktabs}
\usepackage{multirow}
\usepackage{multicol}

\newcommand{\ep}{\epsilon}
\newcommand{\I}{{\cal I}}

\newcommand{\pp}{{\cal P}}

\newcommand{\be}{\begin{equation}}
\newcommand{\ee}{\end{equation}}
\newcommand{\bea}{\begin{eqnarray}}
\newcommand{\eea}{\end{eqnarray}}

\def\renM{{\cal M}}

\def\asc{\bigg( \frac{\hat\alpha_s}{4 \pi} \bigg)}
\def\aec{\bigg( \frac{\hat\alpha}{4 \pi} \bigg)}
\def\asr{\bigg( \frac{\alpha_s}{4 \pi} \bigg)}
\def\aer{\bigg( \frac{\alpha}{4 \pi} \bigg)}

\title{${\mathcal{O}(\alpha_s^2 \alpha)}$ corrections to quark form factor}

\author[a,b]{Tanmoy Pati,}
\author[a,b]{Narayan Rana,}
\author[b,c]{V. Ravindran}

\affiliation[a]{School of Physical Sciences, National Institute of Science Education and Research, Jatni 752050, India}
\affiliation[b]{Homi Bhabha National Institute, Training School Complex, Anushakti Nagar, Mumbai 400094, India}
\affiliation[c]{The Institute of Mathematical Sciences, Taramani, Chennai 600113, India}

\emailAdd{tanmoy.pati@niser.ac.in, narayan.rana@niser.ac.in, ravindra@imsc.res.in}

\abstract{
We present the analytic results for the non-singlet contributions to the three-loop mixed 
strong-electroweak ${\mathcal{O}}(\alpha_s^2\alpha)$ virtual corrections to the quark form factors. 
The primary challenge of this computation arises from the presence of massive vector bosons within the loops. 
This significantly increases the complexity of the integration-by-parts reduction of the scalar integrals 
and complicates their evaluation via the method of differential equations.
To obtain the physical results, we perform the appropriate ultraviolet renormalization and 
subtract the universal infrared divergences. The resulting finite remainders are expressed 
in terms of Harmonic Polylogarithms and Generalized Polylogarithms.
}

\colorlet{shadecolor}{gray!14}

\begin{document}

\tikzset{
photon/.style={decorate, draw=black,
    decoration={coil,aspect=0,segment length=7pt,amplitude=2pt}},    
Zboson/.style={decorate, draw=red,
    decoration={snake,aspect=0,segment length=2pt,amplitude=1pt}},   
Wboson/.style={decorate, draw=red,
    decoration={snake,aspect=0,segment length=3pt,amplitude=1pt}},   
gluon/.style={decorate, draw=black,
    decoration={coil,aspect=0.5,segment length=5pt,amplitude=1pt}},
fermion/.style={draw=black,
      postaction={decorate},decoration={markings,mark=at position .55
        with {\arrow[draw=black]{>}}}},   
scalar/.style={draw=black,
      postaction={decorate},decoration={}},           
vector/.style={decorate, decoration={snake}, draw} 
}

\preprint{~}

\allowdisplaybreaks[4]
\unitlength1cm
\maketitle
\flushbottom


\section{Introduction}

The Drell-Yan (DY) production of a lepton pair~\cite{Drell:1970wh} constitutes one of the benchmark processes for physics investigations at the Large Hadron Collider (LHC). With its substantial production cross-section and distinct experimental signature, DY processes can be measured with minimal experimental uncertainty. This makes them pivotal for stringent tests of the Standard Model (SM), particularly in facilitating the precise determination of key weak sector parameters, such as the W boson mass and the sine of the weak mixing angle.
In addition, the DY process is instrumental in constraining parton distribution functions, calibrating detectors, and determining collider luminosity. Moreover, numerous scenarios involving physics beyond the SM yield final states that closely resemble those of the DY process. Consequently, it serves as a crucial SM background in the pursuit of New Physics.

Given their multifaceted applications, maintaining precise experimental and theoretical control over DY processes is essential to advance future research at particle colliders.
To match the experimental precision, theoretical predictions must be pushed to the highest order in the perturbative expansion in both the strong and electroweak (EW) couplings, $\alpha_s$ and $\alpha$, respectively.
Intensive computational efforts in pure Quantum Chromodynamics (QCD) have established a formidable baseline.
The state-of-the-art began with calculations of the next-to-leading-order~(NLO)~\cite{Altarelli:1979ub}
and next-to-next-to-leading-order~(NNLO)~\cite{Hamberg:1990np,Harlander:2002wh} QCD corrections to the total cross section. These were followed by differential NNLO computations including the leptonic decay of the vector boson~\cite{Anastasiou:2003yy,Anastasiou:2003ds,Melnikov:2006kv,Catani:2009sm,Catani:2010en}.
The next-to-next-to-next-to-leading-order~(N$^3$LO) QCD corrections have been obtained for the inclusive production of a virtual photon~{\cite{Duhr:2020seh,Chen:2021vtu}} and of a $W$ boson~\cite{Duhr:2020sdp}, alongside the computations of fiducial cross sections at this order~\cite{Camarda:2021ict,Chen:2022cgv,Neumann:2022lft,Campbell:2023lcy}.
The NLO EW corrections were obtained
for $Z$ production in refs.~\cite{Baur:2001ze,Zykunov:2005tc,CarloniCalame:2007cd,Arbuzov:2007db,Dittmaier:2009cr},
and
for $W$ production in refs.~\cite{Dittmaier:2001ay,Baur:2004ig,Zykunov:2006yb,Arbuzov:2005dd,CarloniCalame:2006zq}. 
However, as the precision derived from pure QCD calculations has reached this $\text{N}^3\text{LO}$ level, the remaining uncertainties stemming from the EW sector and their interplay with strong interactions become increasingly important. Mixed QCD-EW corrections, hence, became essential for achieving true precision. 

Initial efforts derived mixed QCD-QED corrections for inclusive~\cite{deFlorian:2018wcj} on-shell Z production, later extended to fully differential off-shell Z decay~\cite{Delto:2019ewv,Cieri:2020ikq}. 
Subsequent work presented complete ${\cal O}(\alpha_s \alpha)$ computations for on-shell $Z$ and $W$ boson production~\cite{Bonciani:2016wya,Bonciani:2019nuy,Bonciani:2020tvf,Buccioni:2020cfi,Behring:2020cqi,Bonciani:2021iis}. Next, to obtain results beyond the on-shell approximation, the pole approximation~\cite{Denner:2019vbn} 
was used in refs.~\cite{Dittmaier:2014qza,Dittmaier:2015rxo,Dittmaier:2024row}.
Further steps beyond the pole approximation include the ${\cal O}(n_F \alpha_s \alpha)$ contributions~\cite{Dittmaier:2020vra} to the DY cross section and, 
for the charged-current process, a mixed QCD-EW computation where the two-loop amplitude, initially treated in the pole approximation~\cite{Buonocore:2021rxx}, has now been exactly evaluated~\cite{Armadillo:2024nwk}. 
Finally, the complete computation of mixed QCD-EW corrections for the neutral-current process has been reported for massive~\cite{Bonciani:2021zzf} and massless leptons~\cite{Buccioni:2022kgy}, based on exact two-loop amplitudes~\cite{Armadillo:2022bgm,Bonciani:2016ypc,Heller:2020owb,Heller:2019gkq,Hasan:2020vwn}.
These mixed QCD-EW corrections have been found to be notably larger than initially anticipated in many kinematic regions. This indicates the necessity of pushing theoretical calculations even further by including higher-order mixed corrections, specifically the ${\cal O}(\alpha_s^2 \alpha)$ contributions. Achieving these extremely precise theoretical predictions is essential for fully exploiting the experimental data from modern particle colliders like the LHC. A key component required for obtaining these challenging corrections is the calculation of the quark form factors (FFs) at this specific mixed order. These FFs are fundamental building blocks that encapsulate the virtual corrections necessary for constructing the full higher-order cross section.

The Feynman diagram topologies contributing to these quark FFs can be categorized into three distinct groups based on the EW bosons involved. The first group comprises diagrams with a massless photon in the loop, exhibiting topologies that are subsets of those found in three-loop pure QCD corrections. 
These results were presented in \cite{AH:2019pyp}.
The second group consists of diagrams featuring a single massive Z boson, for which we previously computed the necessary master integrals (MIs) in Ref.~\cite{Pati:2025ivg}. 
The final, third group involves diagrams containing either a single massive $W$ boson or a triple vector boson vertex. The single $W$ boson MIs can be straightforwardly derived from the $Z$ boson MIs via an appropriate change of variables. However, while the topologies featuring the triple vector boson vertex are mostly subsets of those appearing in three-loop mixed QCD-EW corrections to Higgs boson production~\cite{Bonetti:2017ovy}, MIs for some sub-topologies within this last scenario are still missing. 
In this paper, we present the non-singlet contributions of
${\cal O}(\alpha_s^2 \alpha)$ corrections to the quark FFs at three loops, including the computation of these missing MIs for the sub-topologies featuring the triple vector boson vertex.
The singlet contributions are defined as those originating from diagrams with two separate Dirac traces, 
each containing an odd number of $\gamma_5$ matrices. We defer these to a future study.

The structure of this paper is as follows. In Section \ref{sec:theory}, we define the massless quark FFs and establish the theoretical setup, including the ultraviolet (UV) renormalization procedure. Section \ref{sec:comp} details the methodology employed for the calculation. 
Section \ref{sec:result} presents the analytical results for the three-loop virtual mixed QCD-EW corrections, organized by their color and coupling structure. Finally, we conclude in Section \ref{sec:conclusion}.

\section{Theoretical framework}
\label{sec:theory}

\subsection{The scattering process and notations}
The scattering process under study is the production of an off-shell $Z$ boson in quark-antiquark annihilation
\begin{equation}
    q (p_1) + \bar{q} (p_2) \rightarrow Z^* (q) \,.
\end{equation}
$p_1$ and $p_2$ are the momenta of the incoming quark and anti-quark, respectively.
$q$ is the momentum of the $Z$ boson with $q^2 = s$.
The on-shell conditions imply
\begin{equation}
    p_1^2 = 0 \,,\, p_2^2 = 0\,.
\end{equation}
The UV renormalized amplitude of this partonic process admits a perturbative expansion in the two coupling constants,
${\alpha}_s$ and ${\alpha}$, the renormalized strong and electromagnetic coupling constant, respectively,
as in the following
\begin{align}
 |\renM \rangle &=
 \sum_{m,n=0}^{\infty} \asr^m \aer^n |\renM^{(m,n)} \rangle \,.
\end{align}
We consider the three-loop mixed QCD-EW corrections (${\mathcal{O}}(\alpha_s^2 \alpha)$) to the partonic process, that is the coefficient $|\renM^{(2,1)} \rangle$.
The generic amplitude can be written as 
\begin{align}
    |\renM \rangle = -i \delta_{cd} ~ \bar{u}_c (p_1) ~ \Big( F_1 \, \gamma^\mu - F_2 \, \gamma^\mu \gamma_5 \Big) ~ u_d (p_2) \,.
\end{align}
$\bar{u}_c (p_2)$ and $u_d (p_1)$ are the bi-spinors of the incoming anti-quark and quark, respectively.
The ${F}_i$'s also admit perturbative expansion in ${\alpha}_s$ and ${\alpha}$ as
\begin{align}
 {F}_i &= \sum_{m,n=0}^{\infty} \asr^m \aer^n {F}_i^{(m,n)} \,.
\end{align}
For pure QCD corrections, i.e., for $n=0$,
\begin{equation}
    {F}_1^{(m,0)} = {F}^{(m,0)} v_q \,,~~ {F}_2^{(m,0)} = {F}^{(m,0)} a_q \,,
\end{equation}
where,
\begin{equation}
    v_q = \Big( \frac{I_3^q}{2} - s_w^2 Q_q \Big) \,, a_q = \frac{I_3^q}{2} \,.
\end{equation}
$I_3^q$ is the third component of the quark isospin,
$Q_q$ is the electric charge of the quark
and 
$s_w$ is the sine of weak mixing angle.
We define the FFs such that $F^{(0,0)}=1$.
The inclusion of EW vector bosons in the loops alters the chiral structure of the amplitudes. 
In multi-loop calculations, the presence of chiral quantities within dimensional regularization introduces 
the well-known challenge of generalizing the inherently four-dimensional object $\gamma_5$ to $d=4-2\ep$ space-time dimensions. 
For non-singlet contributions, characterized by an even number of $\gamma_5$ within every Dirac trace, 
we employ the naive anti-commutation scheme: 
\begin{equation}
 \{ \gamma_\mu, \gamma_5 \} = 0 \,.
\end{equation}
Conversely, singlet contributions arise from diagrams containing separate Dirac traces, each with an odd number of $\gamma_5$. 
In pure QCD, this distinction is typically between diagrams with open 
fermion lines connected to a chiral vertex (non-singlet) and those where a closed fermion loop is attached 
to the vertex (singlet). However, the presence of internal EW bosons complicates this classification. 
Specifically, certain non-singlet diagrams containing an internal quark loop, and thus two distinct Dirac traces, 
exhibit singlet-type behavior where each trace contains an odd number of $\gamma_5$ matrices.
Due to the complexities, such contributions are treated alongside the singlet diagrams and are deferred to future work. 
In this paper, we focus exclusively on the remaining non-singlet contributions where the naive anti-commutation relation remains consistent.
The chiral structure of the non-singlet FFs  (for $u$-quark) in terms of $v_u, a_u$ up to three loops with $n=1$ can be expressed as
\begin{align}
    F_1^{(m,1)} &= 
       \frac{ v_u Q_u^2 }{c_w s_w}  F_{\gamma,1}^{(m,1)} 
     + \frac{v_u \sum_{f} Q_f^2}{c_w s_w}  F_{\gamma,2}^{(m,1)}  
     + \frac{ v_u (v_u^2 + 3 a_u^2) }{c_w^3 s_w^3}  {F}_{Z,1}^{(m,1)} 
     + \frac{ v_u \sum_f (v_f^2 + a_f^2) }{c_w^3 s_w^3}  {F}_{Z,2}^{(m,1)}
    \nonumber\\
    &+ \frac{(v_{d} + a_{d})}{8 c_w s_w^3} {F}_{W,1}^{(m,1)}  
     + \frac{(v_u + a_u)}{8 c_w s_w^3} {F}_{W,2}^{(m,1)}  
     + \frac{v_u}{8 c_w s_w^3} {F}_{W,3}^{(m,1)}  \,.
%
\\
    F_2^{(m,1)} &= 
       \frac{ a_u Q_u^2 }{c_w s_w} F_{\gamma,1}^{(m,1)} 
     + \frac{a_u \sum_{f} Q_f^2}{c_w s_w} F_{\gamma,2}^{(m,1)}  
     + \frac{ a_u (3 v_u^2 + a_u^2) }{c_w^3 s_w^3}  {F}_{Z,1}^{(m,1)} 
     + \frac{ a_u \sum_f (v_f^2 + a_f^2) }{c_w^3 s_w^3}  {F}_{Z,2}^{(m,1)}
    \nonumber\\
    &+ \frac{(v_{d} + a_{d})}{8 c_w s_w^3} {F}_{W,1}^{(m,1)}  
     + \frac{(v_u + a_u)}{8 c_w s_w^3} {F}_{W,2}^{(m,1)}  
     + \frac{a_u}{8 c_w s_w^3} {F}_{W,3}^{(m,1)}  \,.
\end{align}
Note that ${F}_{Z,2}^{(m,1)}$ and ${F}_{W,3}^{(m,1)}$ receive contributions from diagrams with a similar topology:
a vector boson connected within an internal quark loop. 
For the $Z$-boson case, the individual flavor contributions are distinct and must be explicitly summed.
Conversely, in the $W$-boson case, the contributions from all flavors are identical. Hence, the summation yields a factor of
$\frac{n_F}{2}$ where $n_F$ denotes the total number of quark flavors. 
The three-loop  ${\mathcal{O}}(\alpha_s^2 \alpha)$ corrections to the
matrix element can be written in terms of the FFs as
\begin{align}
  \langle \renM^{(0,0)} |\renM^{(2,1)} \rangle &= 
  \frac{ (v_u^2 + a_u^2)}{c_w^2 s_w^2} \bigg(  Q_u^2 ~ F_{\gamma,1}^{(2,1)} + \big( \sum_{f} Q_f^2 \big) ~ F_{\gamma,2}^{(2,1)} \bigg) 
  \nonumber\\
  &+ \frac{(v_u^4 + 6 v_u^2 a_u^2 + a_u^4)}{c_w^4 s_w^4} ~ {F}_{Z,1}^{(2,1)}
   + \frac{(v_u^2 + a_u^2) \sum_f (v_f^2 + a_f^2)}{c_w^4 s_w^4} ~ {F}_{Z,2}^{(2,1)}
  \nonumber\\
  &+ \frac{(v_u + a_u)(v_{d} + a_{d})}{8 c_w^2 s_w^4} {F}_{W,1}^{(2,1)}  
     + \frac{(v_u + a_u)^2}{8 c_w^2 s_w^4} {F}_{W,2}^{(2,1)}  
     + \frac{(v_u^2 + a_u^2)}{8 c_w^2 s_w^4} {F}_{W,3}^{(2,1)} \,.
\label{eq:ampff}
\end{align}
The results for $F_{\gamma,i}^{(2,1)}$ were previously presented in ref.~\cite{AH:2019pyp}. 
Consequently, the following discussion focuses exclusively on the technical details and methodology 
relevant to the computation of the FFs ${F}_{Z,i}^{(2,1)}$ and ${F}_{W,i}^{(2,1)}$.

\subsection{Ultraviolet renormalization}
\label{sec:uv}

Prediction of the hadronic cross section requires that the bare couplings and masses defined in the Lagrangian be re-expressed in terms of physical parameters through the process of UV renormalization. 
Our computation employs the background field gauge~\cite{Denner:1994xt}, a choice that conveniently organizes the overall calculation into two distinct and separately UV-finite contributions. The first contribution is the combined vertex correction alongside the necessary quark wave function and $\alpha_s$ renormalization. The second contribution comprises the charge renormalization and the renormalization of the external gauge boson wave function. 
The crucial advantage of this decomposition is that isolating the vertex correction along with the quark wave function and $\alpha_s$ renormalization renders that subset of contributions UV finite on its own.
As the focus of this paper is the three-loop virtual contributions to the FFs, we 
consider the first subset and in the following, we will detail the necessary steps. We defer the presentation of the second set of contributions to a future work.

The renormalization of external massless quark wave function receives EW and mixed QCD-EW contributions. 
Due to the presence of massive $Z$ or $W$ bosons, we calculate the constants in the chiral basis to correctly account for the chiral behavior of the counter-terms. 
The left or right chiral constant ($Z_{q,i}$, $i=l$ or $r$) is expanded perturbatively as
\begin{equation}
    Z_{q,i} = 1 + \aec Z_{q,i}^{(0,1)} + \asc \aec Z_{q,i}^{(1,1)} + \asc^2 \aec Z_{q,i}^{(2,1)} + \cdots   \,,
\end{equation}
where,
\begin{align}
 Z_{q,l}^{(m,1)} &= \bigg[ \frac{l_q^2}{s_w^2 c_w^2} \bigg(\frac{\mu^2}{m_Z^2} \bigg)^\ep + \frac{1}{2 s_w^2} \bigg(\frac{\mu^2}{m_W^2} \bigg)^\ep ~ \bigg]
                    z_{q}^{(m,1)}
, \,
 Z_{q,r}^{(m,1)} = \bigg[ \frac{r_q^2}{s_w^2 c_w^2} \bigg(\frac{\mu^2}{m_Z^2} \bigg)^\ep ~ \bigg]
                    z_{q}^{(m,1)}
,
\end{align}
with $l_q = v_q+a_q$ and $r_q=v_q-a_q$.
The constants up to two loops, ($z_{q}^{(0,1)}$ and $z_{q}^{(1,1)}$), were calculated in ref. \cite{Behring:2020cqi}.
For completeness, we present them in the following along with the new, three-loop result $z_{q}^{(2,1)}$ calculated
in the On-Shell (OS) scheme. Consistent with our treatment of the FFs, we have excluded the singlet contributions, 
specifically those originating from diagrams with two separate Dirac traces each containing an odd number of $\gamma_5$ matrices.
\begin{align}
 z_{q}^{(0,1)} &= -\frac{2 (1 - \epsilon) \Gamma(1+\epsilon)}{(2-\epsilon) \epsilon} \,. 
 \nonumber\\
 z_{q}^{(1,1)} &= C_F \bigg[ \frac{(1-3 \epsilon) (3 - 2 \epsilon) \Gamma (1-\epsilon) \Gamma(1+\epsilon) \Gamma(1+2 \epsilon)}{(1-2 \epsilon) (2-\epsilon) \epsilon} \bigg]\,.
 \nonumber\\
 z_{q}^{(2,1)} &=
   \frac{1}{\ep^2} \bigg\{ \frac{11}{3} C_A C_F - \frac{2}{3} C_F n_F \bigg\}
  +\frac{1}{\ep} \bigg\{ \bigg( \frac{11}{2} - 4 \zeta_3 \bigg) C_A C_F + \frac{2}{3} C_F n_F - \frac{3}{2} C_F^2 \bigg\}
  \nonumber\\&
  +\bigg\{ \bigg( -\frac{19}{2} + \frac{55}{2} \zeta_2 - 12 \zeta_3 - \frac{12}{5} \zeta_2^2 \bigg) C_A C_F
  +\bigg(- \frac{3}{2} - 5 \zeta_2  + 4 \zeta_3  \bigg) C_F n_F
  \nonumber\\&
  +\bigg( \frac{89}{4} + 12 \zeta_3 \bigg) C_F^2 \bigg\}
  + {\mathcal{O}} (\ep) \,.
\end{align}
Here, $C_A = N_C$ and $C_F = \frac{N_C^2-1}{2N_C}$ are the adjoint and fundamental Casimir operators, respectively, for SU($N_C$).
$N_C$ denotes the number of colors. 
Beyond the external quark wave function, the strong coupling constant must also be renormalized. 
We adopt the standard $\overline{\text{MS}}$ scheme for $\alpha_s$ renormalization.
However, it is also necessary to include the `weak' effect (${\mathcal{O}}(\alpha \alpha_s)$) 
in the $\alpha_s$ renormalization which was performed in the OS scheme.

\subsection{Universal infrared structure}
\label{sec:ir}
The UV-renormalized FFs still contain infrared (IR) divergences generated by soft and/or collinear massless partons. 
The final-state IR singularities are known to cancel when the virtual amplitude is combined with real emission 
contributions to form an IR-safe observable, as guaranteed by the Kinoshita-Lee-Nauenberg (KLN) theorem. 
Initial-state collinear singularities are absorbed via mass factorization. 
Critically, the IR structure of these FFs is universal.
This universal structure was first successfully characterized for one- and two-loop QCD amplitudes 
by Catani~\cite{Catani:1998bh} and Sterman~\cite{Sterman:2002qn} using universal subtraction operators. 
Subsequently, the factorization of the single pole in terms of soft and collinear anomalous dimensions 
was shown up to two loops \cite{Ravindran:2004mb} and later confirmed at three loops \cite{Moch:2005tm}. 
Catani's proposal has also been generalized beyond two loops in refs.~\cite{Becher:2009cu, Gardi:2009qi}.

The classification of Feynman diagrams is crucial for the systematic analysis of the IR structure.
Following our categorization based on the gauge bosons present in the loops: the first group, 
which exclusively contains the massless photon, exhibits an IR structure analogous to the full three-loop QCD case. 
This is because the photon generates the same type of soft and collinear divergences as the gluon. 
A detailed, dedicated study on the specific IR properties of these diagrams has been presented in ref.~\cite{AH:2019pyp}. 
In contrast, the second and third groups involve the massive electroweak gauge bosons. Since the massive $Z$ and $W$ 
propagators regulate the soft and collinear regions, they do not contribute to the universal IR divergences. 
Consequently, the IR structure of these contributions is significantly simpler, resulting in an effective 
two-loop QCD IR structure.

This distinction simplifies the application of universal subtraction operators, as the complexity level for analyzing the IR structure is reduced.
In the following, we limit our discussion to the second and third groups. 
We follow ref.~\cite{Becher:2009cu} to write the IR structure of the FFs as a multiplicative factor.
This factorization isolates the divergent terms into a factor, $Z_{\text{IR}}$, 
ensuring that the remaining component, $F_{V,i}^{\text{fin}} ({\mu})$ remains finite as $\ep \rightarrow 0$.
\begin{equation}
 F_{V,i} = Z_{\text{IR}} ({\mu}) F_{V,i}^{\text{fin}} ({\mu}) \,.
 \label{eq:ir}
\end{equation}
The factor $Z_{\text{IR}}$ can be determined using the renormalization group equation, with its solution expressed in terms of the anomalous dimension 
and the $\beta$-function.
We present the solution here only up to two loops, as required
\begin{align}
 \ln Z_{\text{IR}}  &= \asr \bigg[ \frac{\Gamma_0'}{4 \ep^2}  + \frac{\Gamma_0}{2 \ep} \bigg]
       + \asr^2 \bigg[ - \frac{3 {\beta}_0 \Gamma_0'}{16 \ep^3} + \frac{\Gamma_1' - 4 {\beta}_0 \Gamma_0}{16 \ep^2} + \frac{\Gamma_1}{4 \ep} \bigg]
       + {\cal O}({\alpha}_s^3) \,,
\end{align}
where, 
\begin{equation}
    \Gamma = - C_F ~ \gamma_{\text{cusp}} \ln \bigg( \frac{\mu^2}{-s} \bigg) + 2 \gamma_q
    \quad \text{and} \quad
    \Gamma' = \frac{\partial}{\partial \ln {\mu}} \Gamma \,.
\end{equation}
$\gamma_{\text{cusp}}$ and $\gamma_{q}$ also are expanded in perturbative series in $\alpha_s$
\begin{equation}
 \gamma_{\text{cusp}} = \sum_{n=1}^\infty \asr^{n} \gamma_{\text{cusp}}^{(n)} \,,
 \quad 
 \textmd{and}
 \quad 
 \gamma_{q} = \sum_{n=1}^\infty \asr^{n} \gamma_{q}^{(n)} \,.
\end{equation}
$\gamma_{\text{cusp}}$ is the massless cusp anomalous dimension and 
$\gamma_{q}$ is the linear combination of collinear and soft anomalous dimension \cite{Ravindran:2004mb}.

\section{Computational details}
\label{sec:comp}
The three-loop FFs were calculated using a conventional computational workflow. 
Feynman diagrams were first generated with \textsc{QGRAF} \cite{Nogueira:1991ex}. 
The \textsc{QGRAF} output was then processed by in-house \textsc{FORM} \cite{Tentyukov:2007mu} routines, 
which converted the diagrams into Feynman amplitudes and managed subsequent manipulations, 
including Dirac, Lorentz, and color algebra. 
Color algebra was performed using the \textsc{FORM} package \textsc{Color} \cite{vanRitbergen:1998pn}.

The total number of diagrams contributing to $F_{Z,i}^{(2,1)}$ is 388, considering a single quark flavor.
For $F_{W,i}^{(2,1)}$, the number of diagrams are 276 and 60, respectively, for a single and double $W$ propagator.
Notably, the $W$-boson contributions necessitated the consideration of two quark flavors due to the charged-current interaction.
The expressions resulting from the algebraic manipulations contain a large number of scalar Feynman integrals. 
We apply the standard method of Integration-By-Parts (IBP) reduction \cite{Tkachov:1981wb,Chetyrkin:1981qh,Laporta:2001dd}. 
This process effectively reduces the 
large initial number of integrals to a significantly smaller, linearly independent set known as Master Integrals (MIs). 
We utilized the public codes \textsc{Kira}~\cite{Maierhoefer:2017hyi, Klappert:2020nbg, Lange:2025fba} 
and \textsc{LiteRed}~\cite{Lee:2012cn,Lee:2013mka} to perform this IBP reduction. 
A crucial step for this procedure is selecting the most suitable integral families. 
In the following sections, we present the details of the integral families, IBP reduction procedure and computation of the remaining 
MIs, that are relevant to the current computation.

\subsection{Integral families}
The 388 Feynman diagrams contributing to $F_{Z,i}^{(2,1)}$ are systematically mapped to the 25 integral families 
presented in ref.~\cite{Pati:2025ivg} using \textsc{Reduze} \cite{Studerus:2009ye,vonManteuffel:2012np}. 
The 276 diagrams involving a single $W$ boson propagator can also be mapped 
to these 25 integral families by choosing the appropriate mass of the propagator. 
The remaining 60 diagrams, those containing the triple vector boson vertex, 
require defining 4 new integral families $P_1, N_1, N_2$ and $N_3$.
\begin{align*}
    & P_{1} : \{
\pp_{1}, \pp_{2}, \pp_{3}-m_W^2, \pp_{1;12}, \pp_{2;12}, \pp_{3;12}-m_W^2, \pp_{32}, \pp_{21}, \pp_{31}, \pp_{1;1}, \pp_{2;1}, \pp_{3;1} \}
\\
    & N_{1} : \{
\pp_{2}, \pp_{3}-m_W^2, \pp_{1;12},\pp_{3;12}-m_W^2, \pp_{32}, \pp_{21}, \pp_{31}, \pp_{1;1}, \pp_{2;1}, \pp_{12;1},\pp_{12-3;12}, \pp_{32;2} \}
\\
    & N_{2} : \{
\pp_{1},\pp_{2}, \pp_{3}-m_W^2, \pp_{1;12},\pp_{2;12},\pp_{3;12}-m_W^2, \pp_{32}, \pp_{21}, \pp_{1;1}, \pp_{3;1}, \pp_{12;1}, \pp_{12-3;12}, \pp_{21;2} \}
\\
    & N_{3} : \{
\pp_{1}, \pp_{3}-m_W^2, \pp_{12},\pp_{23}, \pp_{1,1}, \pp_{12;-2}, \pp_{3;12}-m_W^2, \pp_{13}, \pp_{2;2}, \pp_{3;2},\pp_{2},\pp_{2;12} \}
\end{align*}
where,
\begin{gather*}
    \pp_{i} = k_i^2 \,, ~~~
    \pp_{ij} = (k_i-k_j)^2 \, ~~~
    \pp_{i;j} = (k_i-p_j)^2 \,, ~~
    \pp_{i;jk} = (k_i-p_j-p_k)^2 \,,\\ ~~\pp_{12;1} = (k_1-k_2-p_1)^2 \,,
    \pp_{12-3;12} = (k_1-k_2+k_3-p_1-p_2)^2 \,,\pp_{32;2} = (k_3-k_2-p_2)^2 \,\\
    \pp_{21;2} = (k_2-k_1-p_2)^2 \,, ~~
    \pp_{12;-2} = (k_1-k_2+p_2)^2 \,.
\end{gather*}
We note that the first 3 of these integral families have appeared in the calculation of three-loop mixed QCD-EW corrections to the Higgs boson production \cite{Bonetti:2017ovy}.

\subsection{IBP Reduction}
Contraction of the tensor amplitude with appropriate projectors results in a set of scalar integrals which, 
through the application of IBP reduction techniques, are reduced to a minimal subset of MIs. 
We have performed this reduction in the dot basis where MIs have indices greater than one and no negative indices. 
The IBP reduction has been carried out with the help of \textsc{Kira} and \textsc{LiteRed}. 
The 25 integral families, as presented in ref.~\cite{Pati:2025ivg}, were optimal to compute the MIs.
However, the choice of auxiliary propagators was such that a single family was used to map multiple Feynman integrals. 
This resulted in exceptionally high top-level sectors (reaching values up to $4074$), which pose significant performance challenges for IBP reduction programs.
Furthermore, while most seed integrals were constrained to a $d$-value (total number of propagators) of three, 
a substantial number of top sectors required $s \geq -4$ (where $r, s, d$ follow the \textsc{Kira} convention). 
Reducing these required setting $d=3$, and $s=4$, leading to a massive system of additional equations.
To optimize the process, we strategically reorganized the propagators and expanded the 25 original families into 61, 
effectively mapping high-value top sectors to a maximum value of $1022$. 
Although lowering the top-sector values improved performance, the increased number of distinct topologies 
introduced a new bottleneck: a surge in algebraic IBP and symmetry relations between families. 
This created a dense system of linear equations, particularly slowing down the reduction of families 
at the end of the list, as the reducer must check relations against all preceding families.
To maintain efficiency, we implemented a localized reduction strategy. We manually identified interconnected 
integral families and performed the reduction on these clusters independently.
For diagrams containing the triple vector boson vertex, we identified the following additional topologies.
\begin{center}
\begin{tikzpicture}[scale=0.52]
 \draw[scalar]  (-4,2) -- (-3,1.5);
 \draw[scalar]  (-3,1.5) -- (-2,1);
 \draw[scalar]  (-2,1) -- (-1,0.5);
 \draw[Wboson]  (-1,0.5) -- (0,0);
 \draw[scalar]  (-4,-2) -- (-3,-1.5);
 \draw[scalar]  (-3,-1.5) -- (-2,-1);
 \draw[scalar]  (-2,-1) -- (-1,-0.5);
 \draw[Wboson]  (-1,-0.5) -- (0,0);
 \draw[photon] (0,0) -- (1.5,0); 
 \draw[scalar]  (-1,0.5) -- (-1,-0.5);
 \draw[scalar]  (-2,-1) -- (-2,1); 
 \draw[scalar]  (-3,-1.5) -- (-3,1.5); 
  \node at (1.5,0) {$~$};
 \node at (-0.7,-2.5) {$P_{1}$};
 
\end{tikzpicture}\hspace{0.2cm}\begin{tikzpicture}[scale=0.52]
 \draw[scalar]  (-4,2) -- (-3,1.5);
 \draw[scalar]  (-3,1.5) -- (-2,1);
 \draw[scalar]  (-2,1) -- (-1,0.5);
 \draw[Wboson]  (-1,0.5) -- (0,0);
 \draw[scalar]  (-4,-2) -- (-3,-1.5);
 \draw[scalar]  (-3,-1.5) -- (-2,-1);
 \draw[scalar]  (-2,-1) -- (-1,-0.5);
 \draw[Wboson]  (-1,-0.5) -- (0,0);
 \draw[photon] (0,0) -- (1.5,0);
 \draw[scalar]  (-1,-0.5) -- (-3,1.5); 
 \draw[scalar]  (-3,-1.5) -- (-1,0.5); 
 \draw[scalar]  (-2,-1.) -- (-2,1); 
  \node at (1.5,0) {$~$};
 \node at (-0.7,-2.5) {$N_{1}$}; 
 
\end{tikzpicture}\hspace{0.2cm}\begin{tikzpicture}[scale=0.52]
 \draw[scalar]  (-4,2) -- (-3,1.5);
 \draw[scalar]  (-3,1.5) -- (-2,1);
 \draw[scalar]  (-2,1) -- (-1,0.5);
 \draw[Wboson]  (-1,0.5) -- (0,0);
 \draw[scalar]  (-4,-2) -- (-3,-1.5);
 \draw[scalar]  (-3,-1.5) -- (-2,-1);
 \draw[scalar]  (-2,-1) -- (-1,-0.5);
 \draw[Wboson]  (-1,-0.5) -- (0,0);
 \draw[photon] (0,0) -- (1.5,0); 
 \draw[scalar]  (-1,0.5) -- (-1,-0.5);
 \draw[scalar]  (-2,-1) -- (-3,1.5); 
 \draw[scalar]  (-2,1) -- (-3,-1.5); 
  \node at (1.5,0) {$~$};
 \node at (-0.7,-2.5) {$N_{2}$}; 
 
\end{tikzpicture}\hspace{0.2cm}\begin{tikzpicture}[scale=0.52]
 \draw[scalar]  (-4,2) -- (-1.0,0.5);
 \draw[Wboson]  (0,0) -- (-1.0,0.5);
 \draw[Wboson]  (0,0) -- (-3,-1.5);
 \draw[scalar]  (-4,-2) -- (-3,-1.5);
 \draw[scalar]  (-3,0.5) -- (-1,0.5);
 \draw[photon]  (0,0) -- (1.5,0); 
 \draw[scalar]  (-3,1.5) -- (-3,-0.5);
 \draw[scalar] (-3,-0.5) -- (-2,1);
 \draw[scalar] (-3,-1.5) -- (-3,-0.5);
 \node at (1.5,0) {$~$};
  \node at (1.5,0) {$~$};
 \node at (-0.7,-2.5) {$N_{3}$}; 
\end{tikzpicture}
\end{center}
The first three topologies ($P_1, N_1, N_2$) appear in three-loop mixed QCD-
EW corrections to Higgs boson production \cite{Bonetti:2017ovy}.
while the fourth topology ($N_3$) is specific to our process. 
Since the MIs documented in ref.~\cite{Bonetti:2017ovy} involve high-value top sectors that lead to computationally expensive IBP reductions, we applied our localized reduction strategy to these cases as well. This allowed us to perform the reduction efficiently and subsequently map our results to those MIs.

\subsection{Computation of the remaining Master Integrals}
As previously noted, the MIs for Feynman diagrams involving a single massive boson were computed in ref.~\cite{Pati:2025ivg} 
using the method of differential equations \cite{Kotikov:1990kg,Remiddi:1997ny,Gehrmann:1999as,Argeri:2007up,Henn:2013pwa,Henn:2014qga,Ablinger:2015tua,Ablinger:2018zwz}, 
with results expressed in terms of harmonic polylogarithms (HPLs) or generalized harmonic polylogarithms (GPLs).
Because our basis of MIs is not canonical, some of the individual integrals exhibit contributions of transcendental weight 7 
at the finite $\ep^0$ order of the amplitude. While these higher-weight terms ultimately cancel to leave a maximal weight of 6 
in the final expression, a standard approach would require the explicit evaluation of all such MIs up to the necessary order in $\ep$.
However, evaluating these voluminous higher-weight MIs is extremely challenging due to the complexity arising from multiple base transformations. 
To bypass this, we adopted a strategy that avoids direct evaluation. We identified specific linear combinations of these MIs that 
satisfy a first-order differential equation for the leading power in $\ep$. While this differential equation holds for each single 
power of $\ep$, the non-homogeneous part contains contributions from lower-order terms. 
This approach allows us to extract the weight-6 contributions directly and significantly reduces the computational cost 
by bypassing the need for higher-weight evaluations. The corresponding linear combinations (denoted by ${\mathcal{R}}$) are presented below.
\begin{align}
 {\mathcal{R}}_{5,1} &= 2(1 + 4x - x^2)\mathcal{I}_{5,9} + (1 - x)^2 (1 - 3x)\mathcal{I}_{5,10} + 4 x (1 - x^2)\mathcal{I}_{5,11} \,.
 \\
 {\mathcal{R}}_{5,2} &= 2 x \mathcal{I}_{5,26} - (1 - x)x\mathcal{I}_{5,28} \,.
 \\
 {\mathcal{R}}_{7,1} &= 2 (1 - x) \mathcal{I}_{7,11} + 2 (1 - x) x \mathcal{I}_{7,12} + (1 - 4 x - 3 x^2) \mathcal{I}_{7,13} \,.
 \\
 {\mathcal{R}}_{9,1} &= 2 (1 - x) \mathcal{I}_{9,7} + 2 (1 - x) x \mathcal{I}_{9,8} + (1 - 4 x - 3 x^2) \mathcal{I}_{9,9} \,.
 \\
 {\mathcal{R}}_{9,2} &= 2 (1 + 3 x - x^2) \mathcal{I}_{9,15} + (1 - x)^2 \mathcal{I}_{9,16} + x (3 + 4 x - x^2) \mathcal{I}_{9,17}  \,.
 \\
 {\mathcal{R}}_{11,1} &= (2 + 4 x) \mathcal{I}_{11,2} - (1 - x) (1 - 3 x) \mathcal{I}_{11,3} + (1 - x^2) \mathcal{I}_{11,4} \,.
 \\
 {\mathcal{R}}_{17,1} &= 2 (1 + 3 x - x^2) \mathcal{I}_{17,3} + (1 - x)^2 \mathcal{I}_{17,4} + x (3 + 4 x - x^2) \mathcal{I}_{17,5} \,.
 \\
 {\mathcal{R}}_{20,1} &= \mathcal{I}_{20,1} + (1 - x^2) \mathcal{I}_{20,2} + (1 - x)^2 \mathcal{I}_{20,3} \,.
\end{align}
All MIs originating from topologies $P_1, N_1, N_2$ for diagrams featuring the triple vector boson vertex 
are documented in ref.~\cite{Bonetti:2017ovy}. However, due to the presence of additional sectors in our specific process, 
we identified three new MIs within these topologies. Furthermore, topology $N_3$ is entirely unique to our calculation 
and contributes one additional MI. We have evaluated these four supplemental MIs up to transcendental weight 6
\begin{align}
    {\I}_{P_1,1} &= {P_1}[0, 1, 1, 0, 0, 1, 1, 1, 0, 1, 0, 1],\qquad {\I}_{P_1,2} = {P_1}[1, 0, 1, 0, 0, 1, 1, 1, 0, 0, 0, 1],
    \nonumber\\
    {\I}_{P_1,3} &= {P_1}[1, 1, 1, 0, 0, 1, 1, 0, 1, 0, 0, 1], \qquad {\I}_{N_3,1} = {N_3}[1, 1, 1, 1, 0, 1, 1, 1, 1, 1, 0, 0] \,.
    \label{eq:mis}
\end{align}
To evaluate these MIs, we introduce the variables $y$ as follows
\begin{equation}
 -\frac{s}{m_W^2} = \frac{(1-y)^2}{y} \,.
\end{equation}

\vspace{-0.05cm}

\begin{figure}[h]
 \centering
 \includegraphics[width=0.9\textwidth]{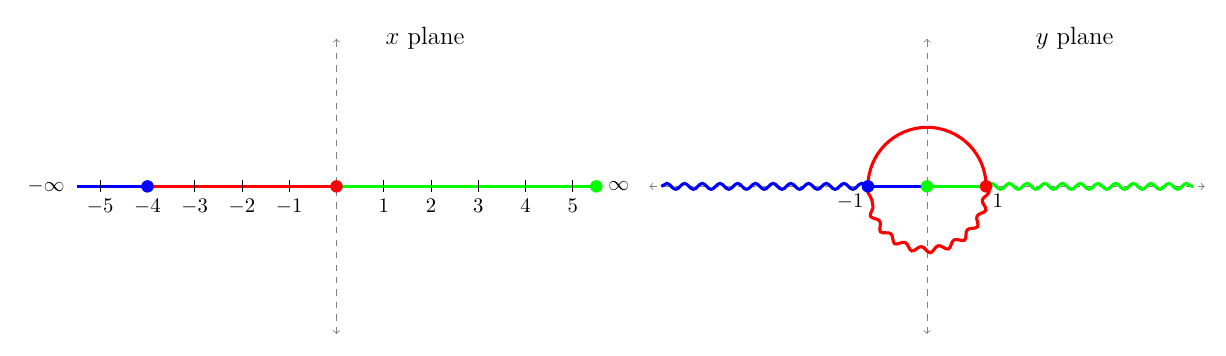}
 \caption{The figure illustrates the transformation between $x$ and $y$ . The left and right panel displays the $x$-plane and the complex $y$-plane, respectively. Colored lines show the mapping of intervals. In the $y$-plane, straight and wiggly lines represent two distinct roots.}
 \label{fig:vary}
\end{figure}

\noindent
In terms of $s$ with prescription $(s+i0^+)$, the two different roots of $y$ can be written as
\begin{equation}
    y^{\pm}=\frac{\sqrt{-4m_W^2+s+i0^+}\pm\sqrt{s+i0^+}}{\sqrt{-4m_W^2+s+i0^+}\mp\sqrt{s+i0^+}} \,.
\end{equation}
In Fig.~\ref{fig:vary}, we illustrate the correspondence between the $x$-plane roots and the complex $y$-plane roots.
Due to our chosen Feynman prescription, the green and blue lines in the $y$ plane lie infinitesimally above and below the real axis, 
respectively. 
These MIs were computed following the same methodology detailed in ref.~\cite{Pati:2025ivg}. 
Various intermediate steps of the calculation utilized the packages 
{\textsc{HarmonicSums}} \cite{Ablinger:2010kw,Ablinger:2011te,Ablinger:2014rba} and 
\textsc{PolyLogTools} \cite{Duhr:2019tlz}. 
The analytic expressions for these four MIs are provided in terms of GPLs in a separate ancillary file. 
We also provide their numerical evaluation at the kinematic point $x= \frac{1}{11}$ below, 
using GiNaC \cite{Bauer:2000cp} for the evaluation of the GPLs. Furthermore, we numerically evaluated 
all contributing MIs, including the ones from ref.~\cite{Bonetti:2017ovy}, across several values of 
$x$ using AMFlow \cite{Liu:2022chg}, finding perfect agreement with the evaluation of our analytic expressions using \textsc{GiNac}.

{\footnotesize
\begin{longtable}{cccccccc}
                    & $\ep^0$  & $\ep^1$ & $\ep^2$ & $\ep^3$ & $\ep^4$ & $\ep^5$ & $\ep^6$ \\ 
                    \hline\\[-1.6ex]
 $\ep^4  P_{1,1}$ & 0.0 & 0.0 & -0.496257 & -2.953627 & -18.363948 & -73.850087 & -309.903083 \\
 $\ep^3  P_{1,2}$ & 0.0 & -0.0833333 & -0.376871 & -2.177518 & -6.518045 & -24.796959 & -57.666901 \\
 $\ep^4  P_{1,3}$ & 0.0 & 0.0 & -0.992515 & -4.914739 & -28.791920 & -102.022561 & -410.434375 \\
 $\ep^6  N_{3,1}$ & 0.0 & 0.0 & 0.0 & -0.638625 & -2.197480 & -9.012971 & -10.936069 \\
\bottomrule 
\end{longtable}

}

\subsection{Construction of the Form Factors}
Once the MIs are obtained, we assemble the non-singlet contributions of the three-loop ${{\mathcal{O}}(\alpha_s^2\alpha)}$ FFs 
by substituting the analytic expressions of the MIs into the IBP-reduced amplitudes. The FFs are then expanded in 
powers of the dimensional regulator $\ep$, where the resulting coefficients are expressed in terms of HPLs and GPLs. 
Directly substituting the IBP reduction rules and MIs into the total amplitude generates a massive intermediate expression.
The resulting expression often exceeds available memory, making the subsequent extraction of 
$\ep$-coefficients computationally challenging.
To manage the algebraic complexity, we perform the simplification on a diagram-by-diagram basis before summing 
the individual contributions at each order in $\ep$. After series expansion, the unsimplified expressions reach 
approximately 60 GB for the $Z$-boson case and 65 GB for the $W$-boson case.
Subsequently, we perform UV renormalization according to the procedure detailed in Section~\ref{sec:uv}. 
Following UV renormalization, we verify that the remaining IR divergences conform to the universal pole structure 
discussed in Section~\ref{sec:ir}. Once the universal IR poles are subtracted, the FFs yield a finite remnant 
of maximal transcendental weight six.

\section{Results}
\label{sec:result}
In this section, we present the results of our calculation. 
We have computed the three-loop non-singlet contributions at ${\mathcal{O}}(\alpha_s^2\alpha)$ 
to the quark FFs. We have also obtained the necessary lower-order contributions: 
the two-loop ${\mathcal{O}}(\alpha_s^2)$ and ${\mathcal{O}}(\alpha_s \alpha)$ corrections expanded up to ${\mathcal{O}}(\ep^2)$, 
and the one-loop ${\mathcal{O}}(\alpha_s)$ and ${\mathcal{O}}(\alpha)$ corrections expanded up to ${\mathcal{O}}(\ep^4)$.
We provide the hard finite remainders, $F_{V,i}^{\text{fin}}$, which also admit a perturbative expansion 
in the strong and electroweak coupling constants,
$\alpha_s$ and $\alpha$:
\begin{equation}
 F_{V,i}^{\text{fin}} = \sum_{m,n=0}^\infty \asr^m \aer^n F_{V,i}^{(m,n),\text{fin}} \,.
\end{equation}
Due to the significant length of the analytic expressions, we present their numerical evaluation over 
a range of center-of-mass (COM) energies, $50$ GeV $\leq \sqrt{s} \leq 150$ GeV. 
The numerical evaluation of the GPLs was performed using \textsc{GiNaC}. 
We set the renormalization scale to the COM energy, $\mu_R^2 = s$, and adopt the following 
values for the vector boson masses
    $m_W = 80.369$ GeV,  $m_Z = 91.1876$ GeV.

\begin{figure}[h]
    \centering
    \includegraphics[width=0.47\linewidth]{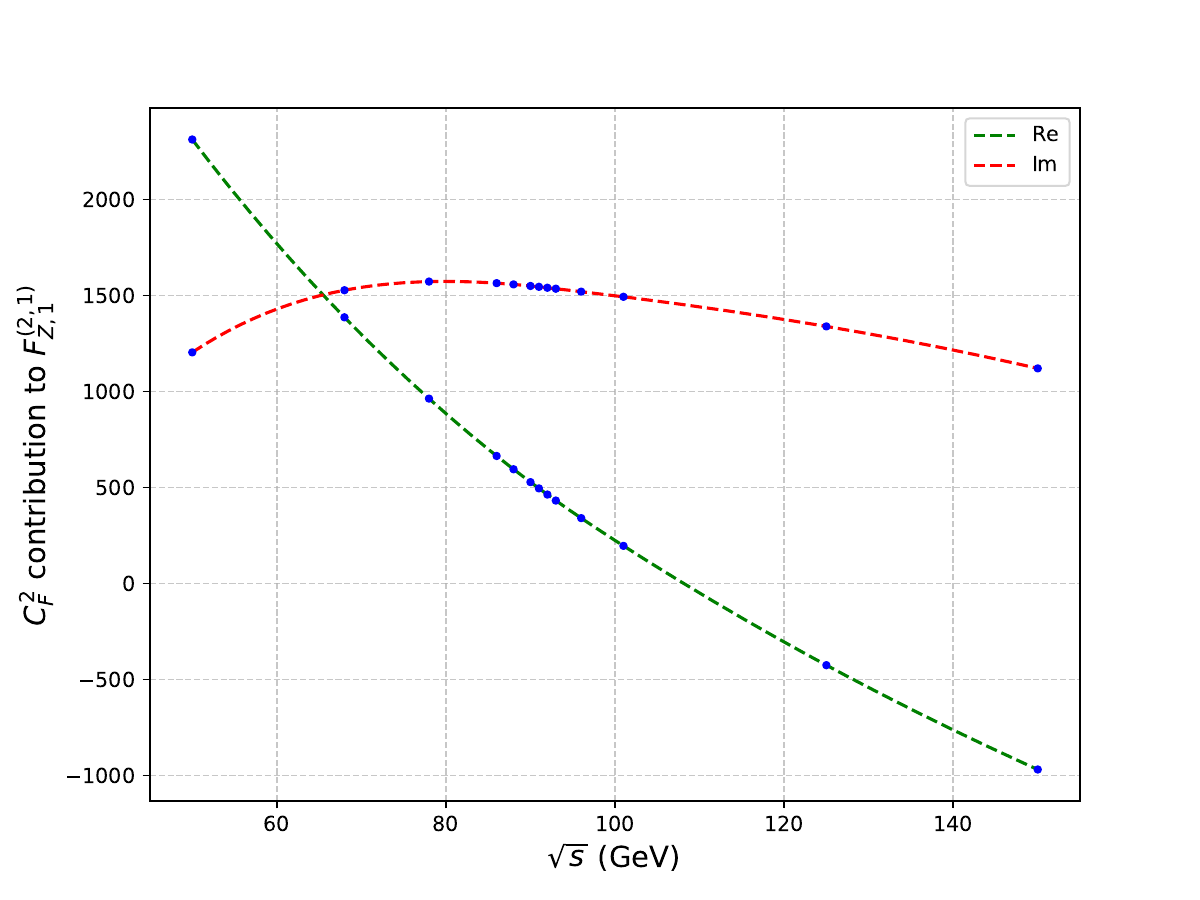}
    \includegraphics[width=0.47\linewidth]{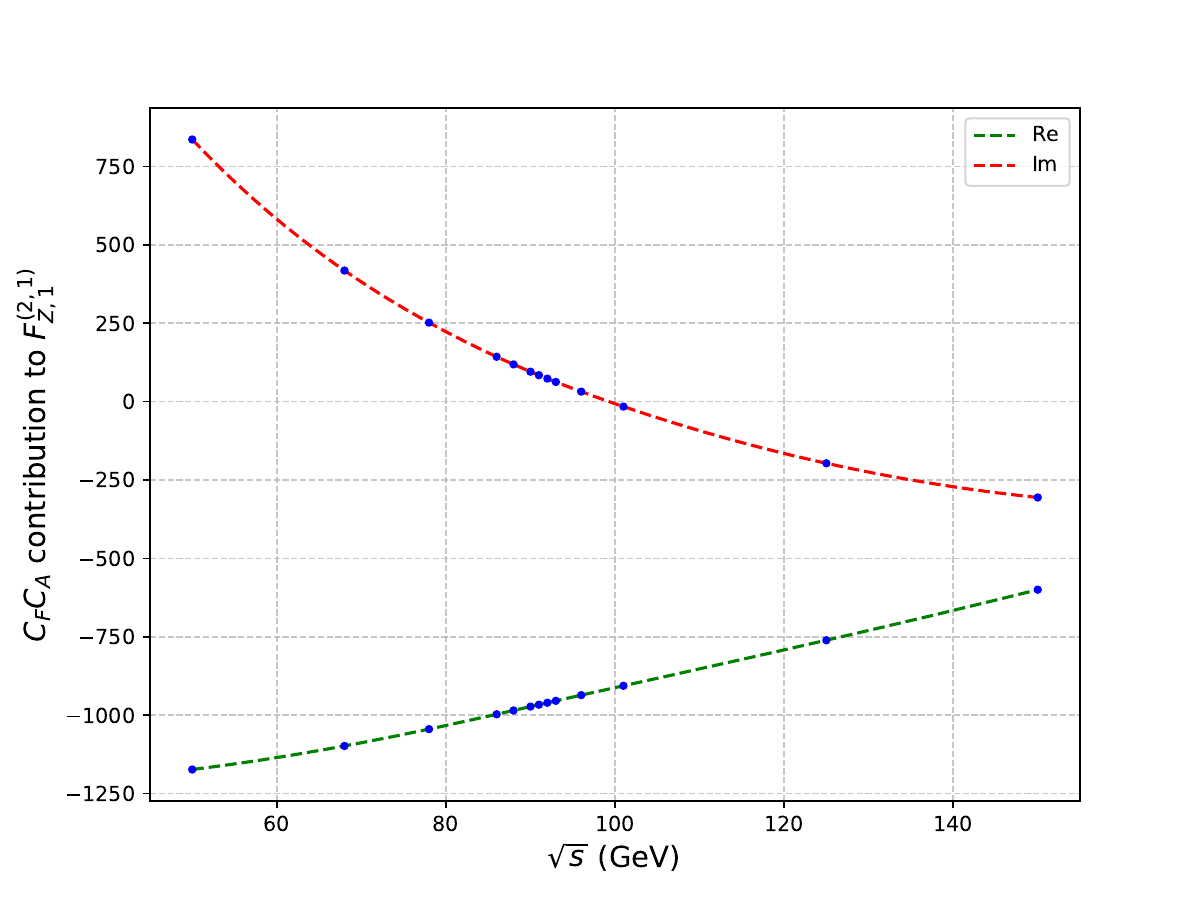}
    \includegraphics[width=0.47\linewidth]{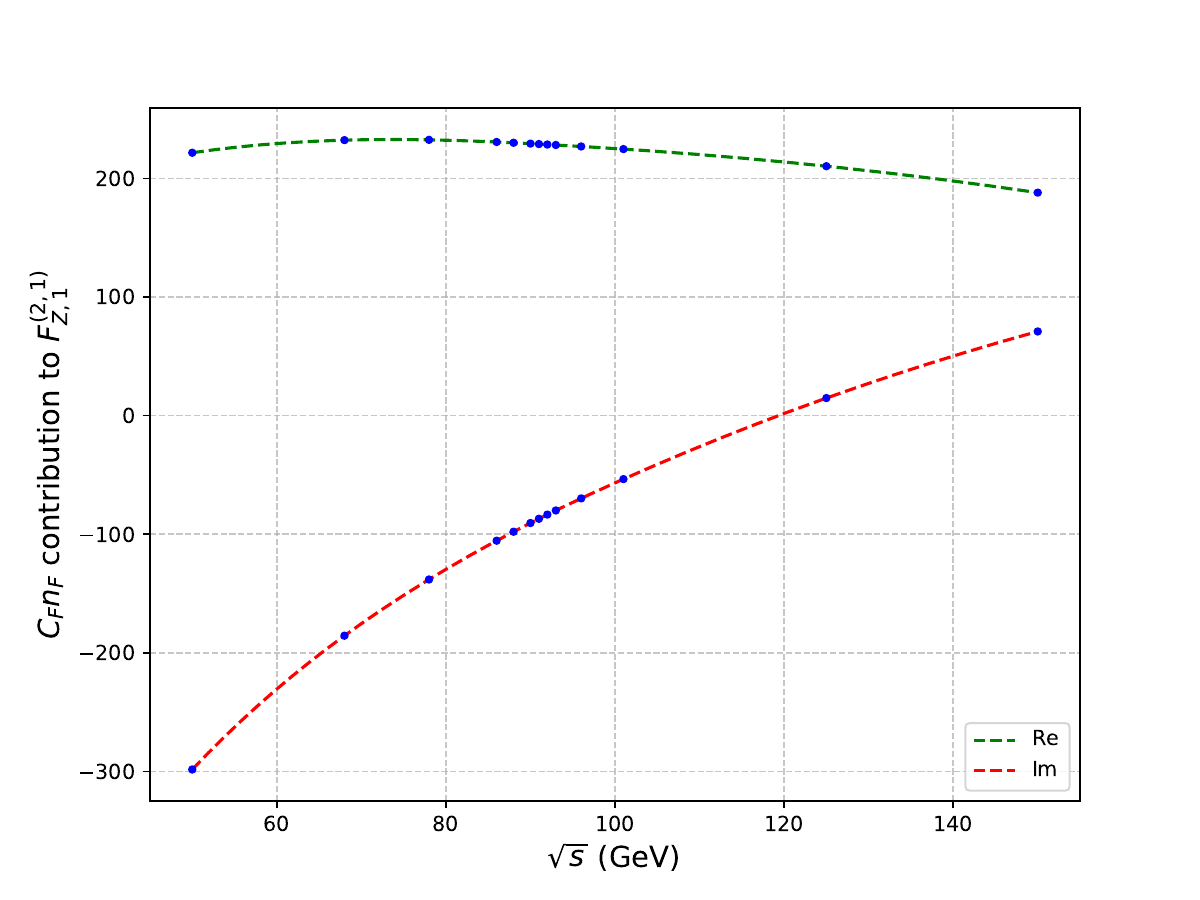}
    \includegraphics[width=0.47\linewidth]{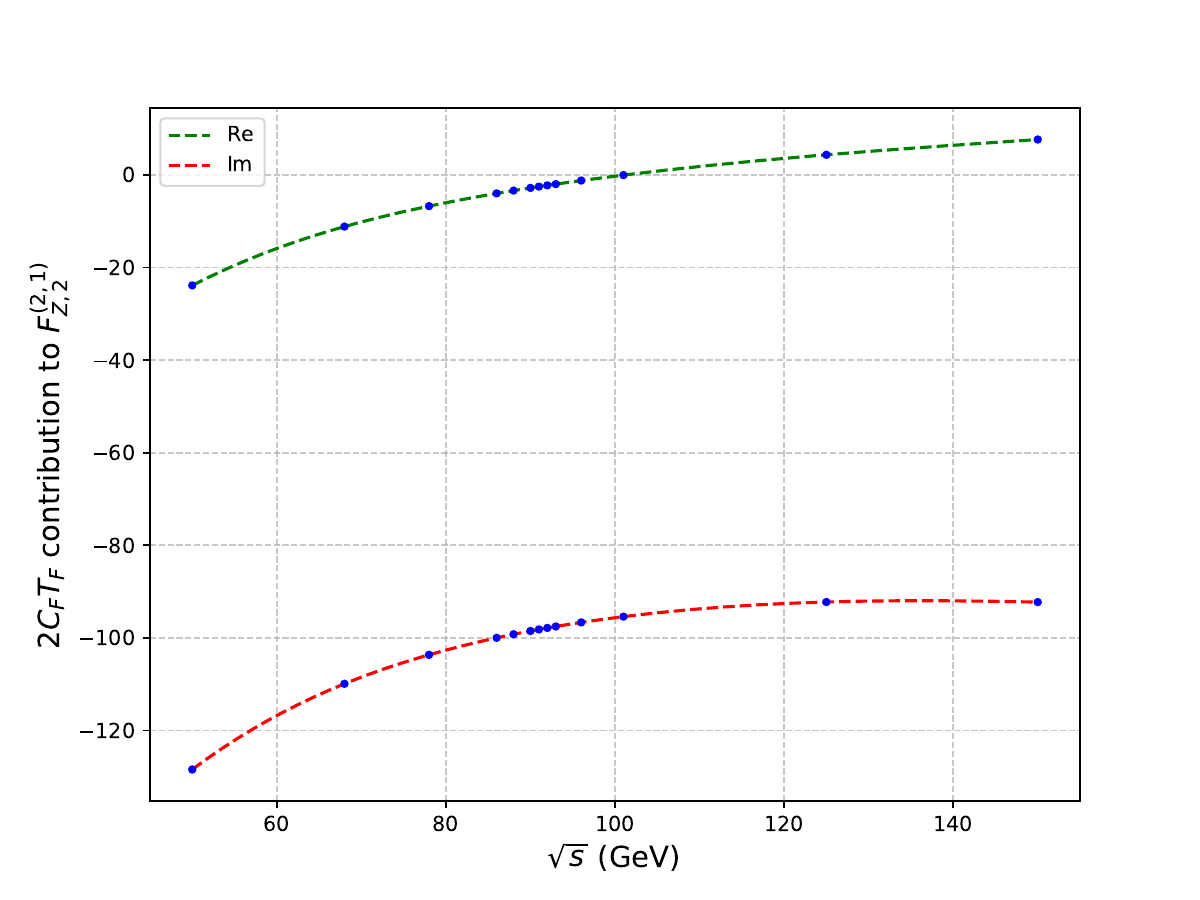}
    \caption{The figure illustrates the individual contributions of each color factor to the coefficients $F_{Z,1}^{(2,1),\text{fin}}$ and $F_{Z,2}^{(2,1),\text{fin}}$ over a range of center-of-mass energies, $50$ GeV $\leq \sqrt{s} \leq 150$ GeV.}
    \label{fig:F21Z}
\end{figure}

The individual contributions from each color factor for the coefficients
$F_{Z,1}^{(2,1),\text{fin}}$ and $F_{Z,2}^{(2,1),\text{fin}}$ are illustrated in Fig.~\ref{fig:F21Z}.
%

\begin{figure}[h]
    \centering
    \includegraphics[width=0.46\linewidth]{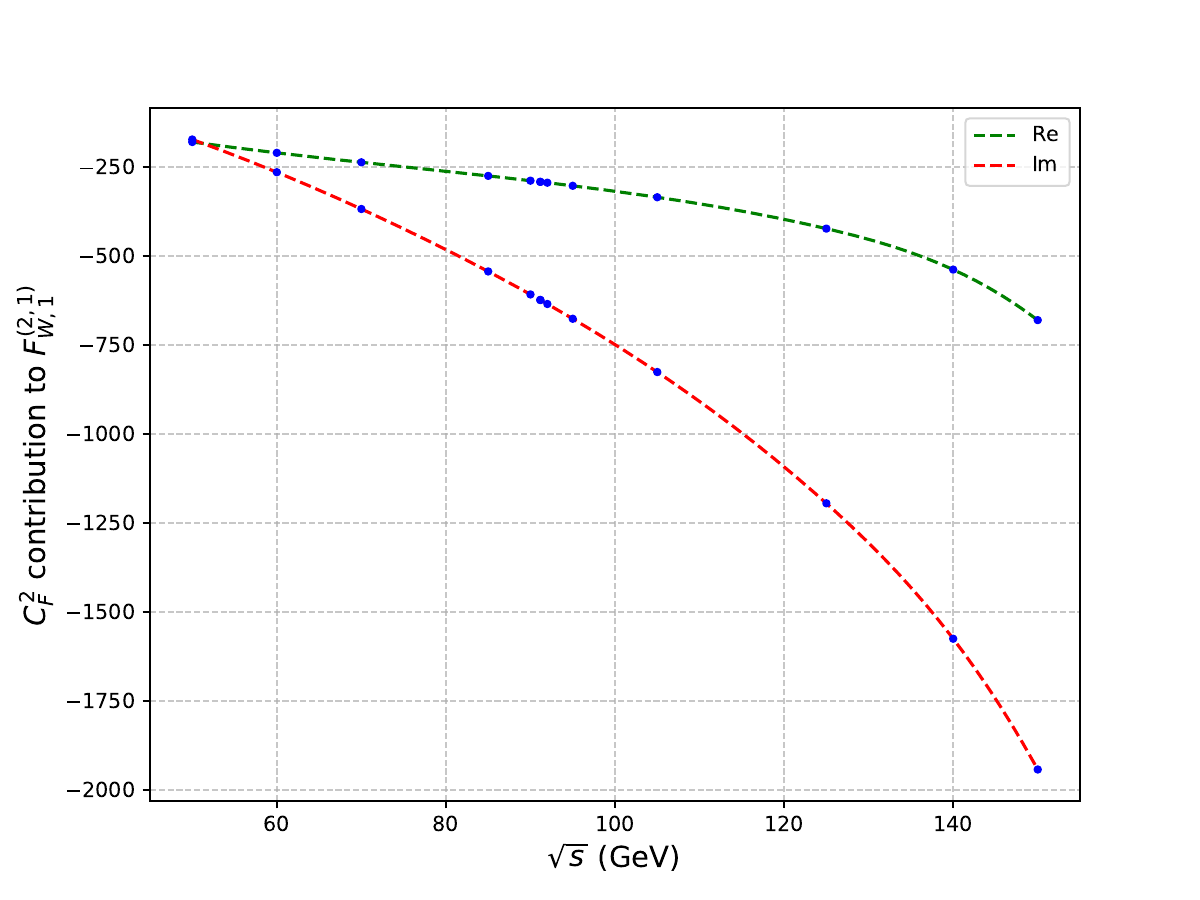}
    \includegraphics[width=0.46\linewidth]{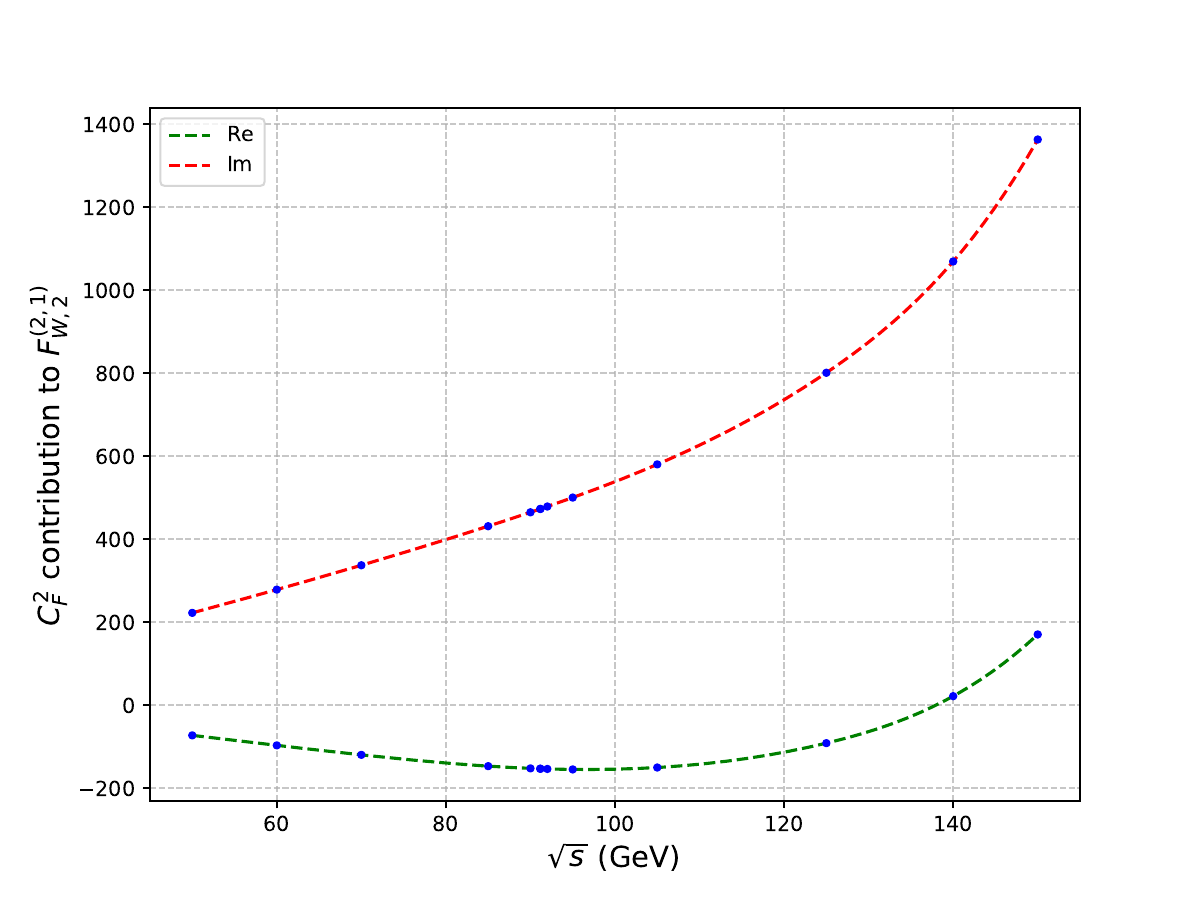}
    \includegraphics[width=0.46\linewidth]{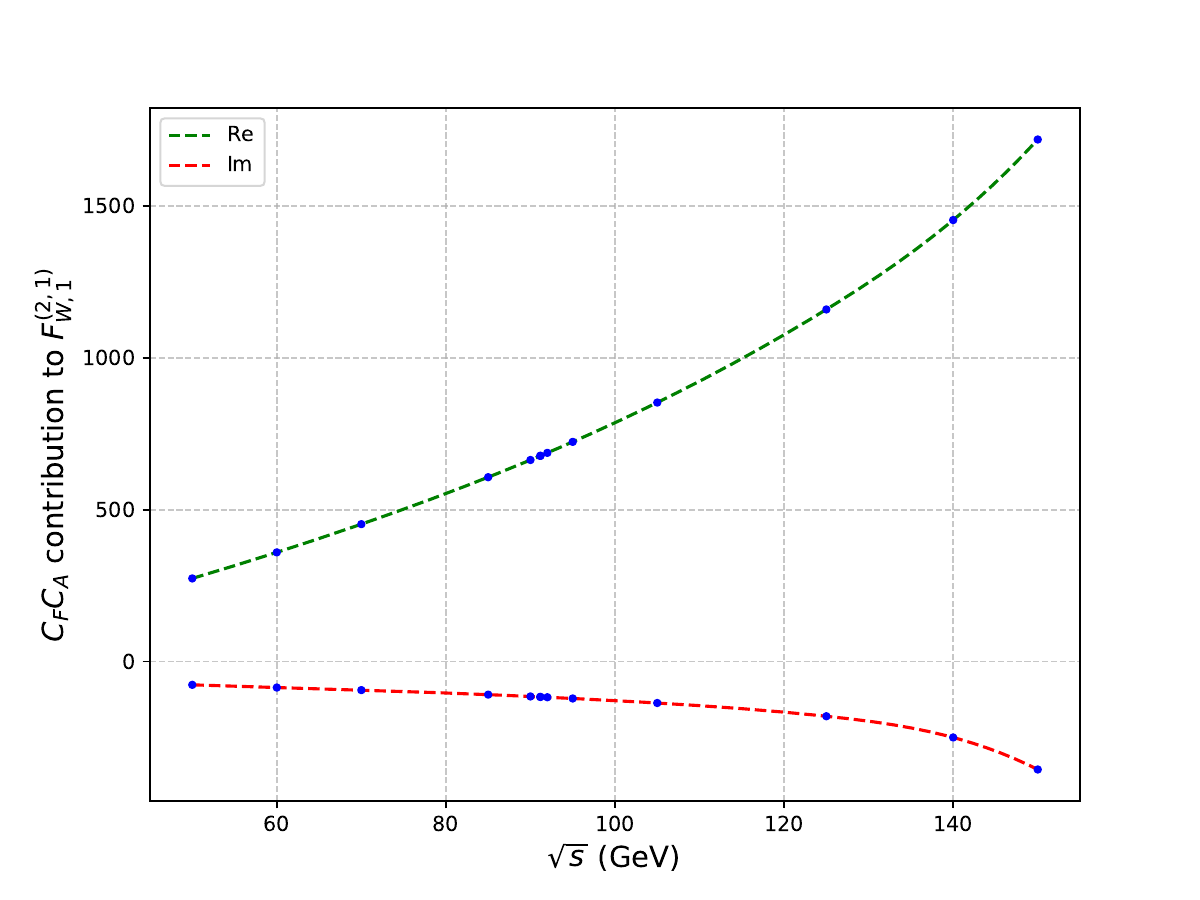}
    \includegraphics[width=0.46\linewidth]{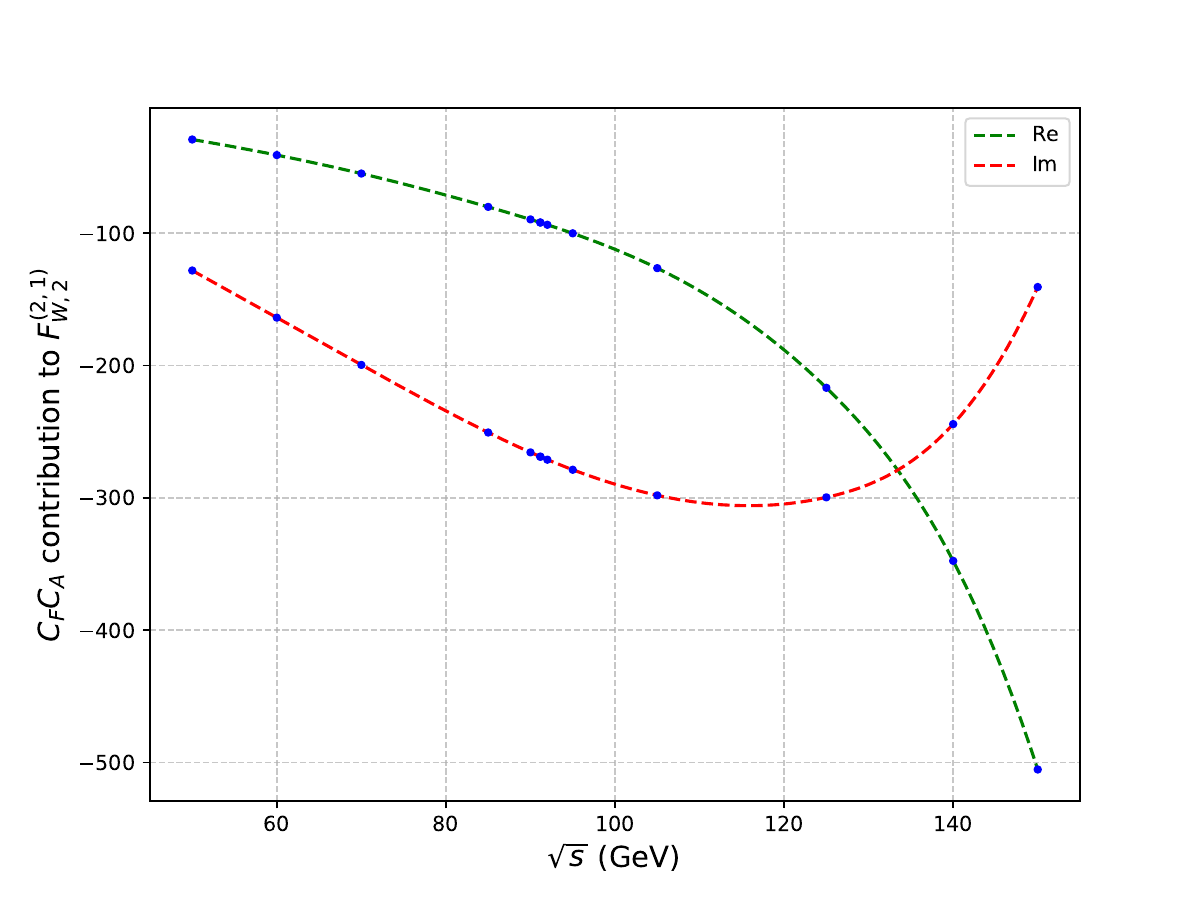}
    \includegraphics[width=0.46\linewidth]{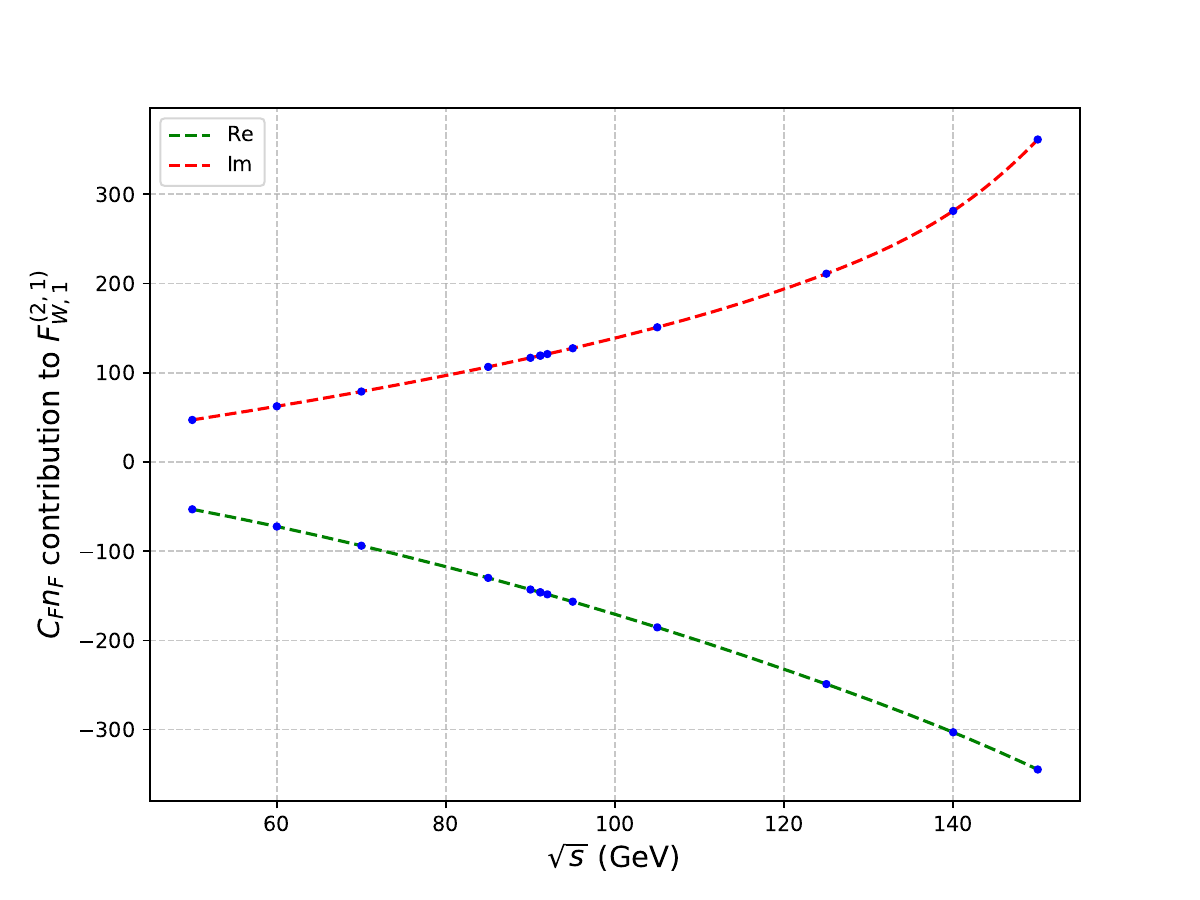}
    \includegraphics[width=0.46\linewidth]{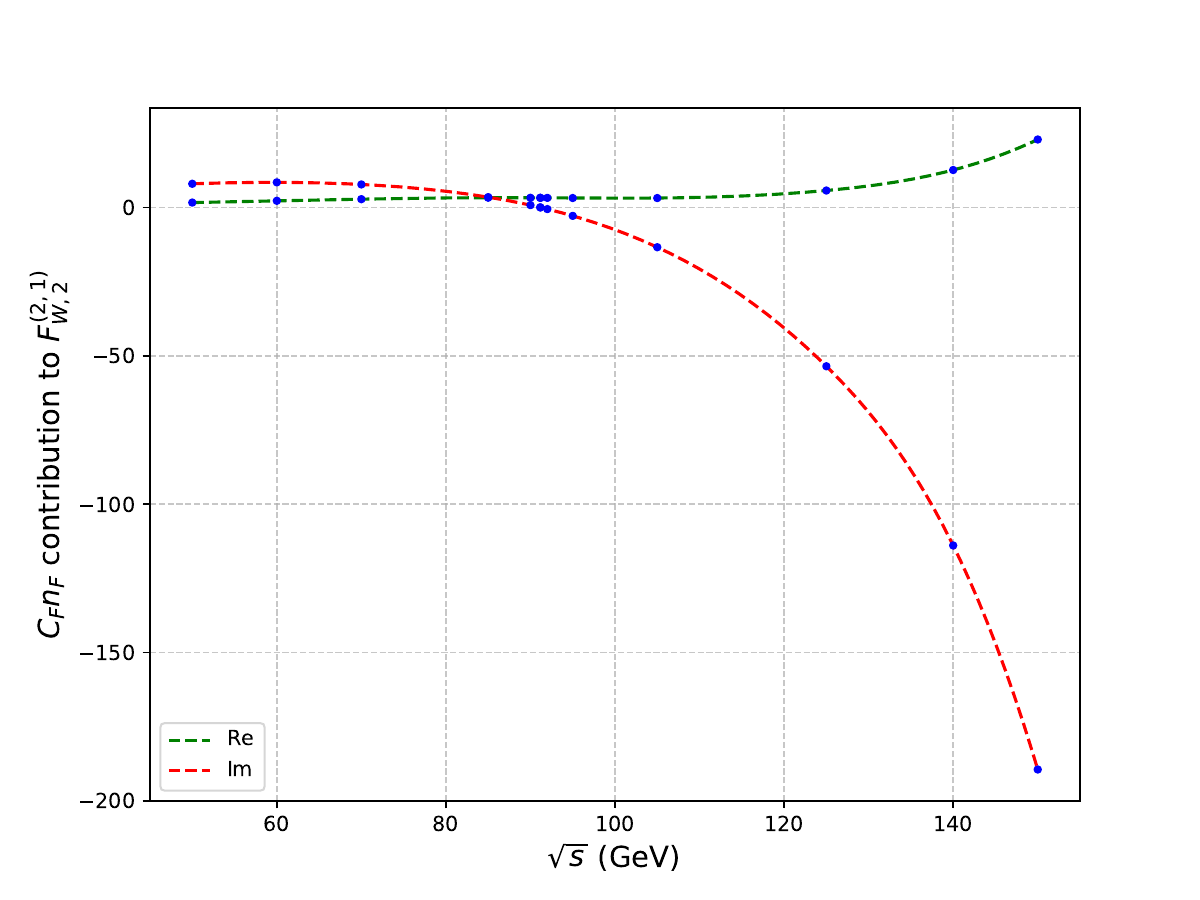}    
    \caption{The figure illustrates the individual contributions of each color factor to the coefficients $F_{W,1}^{(2,1),\text{fin}}$ and $F_{W,2}^{(2,1),\text{fin}}$ over a range of center-of-mass energies, $50$ GeV $\leq \sqrt{s} \leq 150$ GeV.}
    \label{fig:F21W1}
\end{figure}

\begin{figure}[h]
    \centering
    \includegraphics[width=0.48\linewidth]{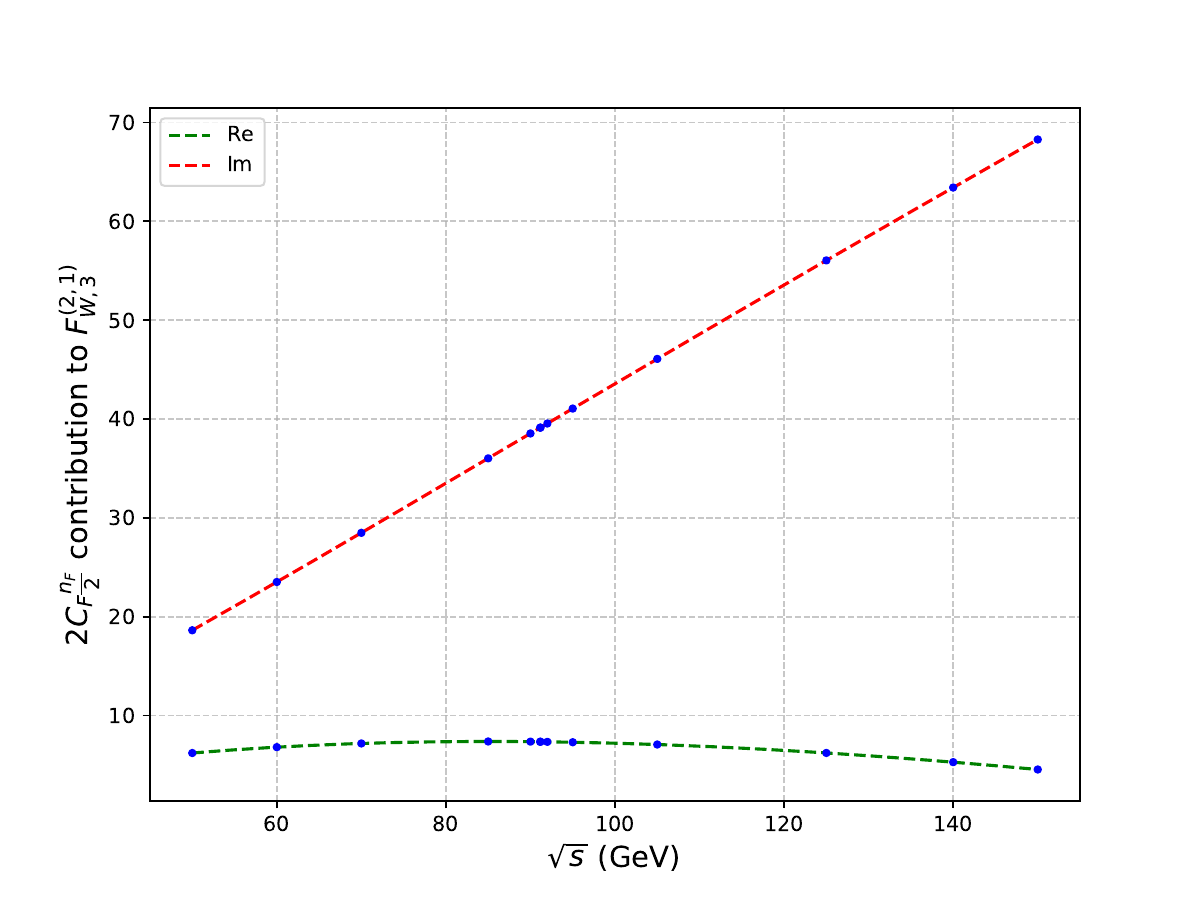}
    \caption{The figure illustrates the $C_F n_F$ contributions to the coefficients $F_{W,3}^{(2,1),\text{fin}}$ over a range of center-of-mass energies, $50$ GeV $\leq \sqrt{s} \leq 150$ GeV.}
    \label{fig:F21W2}
\end{figure}

In a similar manner, Fig.~\ref{fig:F21W1} and Fig.~\ref{fig:F21W2} illustrate the individual contributions of each color factor 
to the coefficients $F_{W,1}^{(2,1),\text{fin}}$, $F_{W,2}^{(2,1),\text{fin}}$ and $F_{W,3}^{(2,1),\text{fin}}$.
Furthermore, we have included an ancillary file, \texttt{result.nb}, with the arXiv submission of this manuscript. 
This file contains the analytic expressions for the finite remainders of the FFs in a \textsc{Mathematica}-readable format.

These FFs serve as essential building blocks for calculating the total ${\mathcal{O}}(\alpha_s^2\alpha)$ 
cross-sections for inclusive production of the $Z$ boson at hadron colliders, such as the LHC. 
To facilitate their application in precision phenomenology, we provide the numerical evaluations 
of the finite remainders, $F_{Z,i}^{(2,1),\text{fin}}$ and $F_{W,i}^{(2,1),\text{fin}}$, in the OS limit $s = m_Z^2$. 
In Table \ref{tab:osffs}, these results are presented. 
We set the renormalization scale $\mu_R^2 = m_Z^2$.
\begin{table}[h!]
    \centering
    \begin{tabular}{c|cc|cc|cc} 
        \toprule
        \multirow{2}{*}{ } & \multicolumn{2}{c|}{ $C_F^2$ } & \multicolumn{2}{c|}{ $C_F C_A$ } & \multicolumn{2}{c}{ $C_F n_F$ } \\
        \cmidrule(lr){2-3} \cmidrule(lr){4-5} \cmidrule(lr){6-7} 
        & Re & Im & Re & Im & Re & Im \\
        \midrule
        ${F}_{Z,1}^{(2,1),fin}$ & 488.632 & 1543.675 & -965.467 & 82.116 & 228.956 & -86.297 \\
        ${F}_{W,1}^{(2,1),fin}$ & -291.776 & -623.940 & 677.526 & -116.207 & -146.197 & 119.198 \\
        ${F}_{W,2}^{(2,1),fin}$ & -152.863 & 473.247 & -92.013 & -268.953 & 3.305 & 0.069 \\
        ${F}_{W,3}^{(2,1),fin}$ &   &  &  &  & 7.354 & 39.138 \\
        \bottomrule
    \end{tabular}
    \caption{Numerical values of the on-shell form factors for $\mu_R^2 = m_Z^2$.}
    \label{tab:osffs}
\end{table}
We note that the diagrammatic contributions to ${F}_{Z,2}^{(2,1),fin}$ include diagrams 
featuring a $Z$-boson coupled to an internal fermion loop. The total contribution from all massless quark 
flavors is proportional to the sum of the squares of the vector and axial-vector couplings, ($v_f^2+a_f^2$). 
This contribution is given by
\begin{align}
 \text{Re}[{F}_{Z,2}^{(2,1),fin}] &= -2.4457 ~ C_F \,, \qquad   
 \text{Im}[{F}_{Z,2}^{(2,1),fin}] = - 98.1178 ~ C_F \,.   
\end{align}

\subsection{Description for the ancillary files}
The ancillary file \texttt{result.nb}, provided in \textsc{Mathematica} format, 
contains the analytic expressions for all FFs listed in Eq.~\ref{eq:ampff}. 
We have defined the following variables
\begin{equation}\label{eq1}
 -\frac{s}{m_Z^2} = x = \frac{(1+x_l)^2}{x_l} \,, \qquad
 -\frac{s}{m_W^2} = x_W = \frac{(1+x_{W,l})^2}{x_{W,l}} = \frac{(1-y)^2}{y}\,.
\end{equation}
In this ancillary file, we adopt the following naming conventions. 
\begin{gather*}
 \text{\texttt{FZ121fin}} = {F}_{Z,1}^{(2,1),fin}, \quad
 \text{\texttt{FZ221fin}} = {F}_{Z,2}^{(2,1),fin},
 \\
 \text{\texttt{FW121fin}} = {F}_{W,1}^{(2,1),fin}, \quad
 \text{\texttt{FW221fin}} = {F}_{W,2}^{(2,1),fin}, \quad 
 \text{\texttt{FW321fin}} = {F}_{W,3}^{(2,1),fin},
 \\
 \text{\texttt{Cf}} = C_F, \quad \text{\texttt{Ca}} = C_A, \quad 
 \text{\texttt{nf}} = n_F, \quad \text{\texttt{Log[muX]}} = \ln (-\mu_R^2/s) \,,
 \\
 \text{\texttt{xL}} = x_l, \quad
 \text{\texttt{xW}} = x_W, \quad
 \text{\texttt{xWL}} = x_{W,l}, \quad 
 \text{\texttt{z2}} = \zeta_2, \quad
 \text{\texttt{z3}} = \zeta_3, \quad
 \text{\texttt{z5}} = \zeta_5, \quad
 \text{\texttt{ln2}} = \ln(2). 
\end{gather*}
The second ancillary file, \texttt{integral.m}, contains the analytic expressions for the four remaining MIs
listed in Eq.~\ref{eq:mis}. 
The data is provided in a \textsc{Mathematica} replacement list format denoting the MIs in \textsc{LiteRed} notation \texttt{j[family,indices][variable]}.

\section{Conclusions}
\label{sec:conclusion}
The DY process continues to play a central role in the precision physics program of the LHC.  
The remarkable experimental accuracy achieved in measurements of EW observables demands theoretical 
predictions of commensurate precision. While significant progress has been made in pure QCD calculations, 
including N$^3$LO results, the relative importance of EW effects and their interplay
with strong interactions has become increasingly pronounced.
In particular, the mixed QCD-EW corrections are no longer a
subleading but an essential ingredient for reliable phenomenology.
Motivated by this, we have computed the three-loop ${\mathcal{O}}(\alpha_s^2 \alpha)$ contributions to
the quark FFs, specifically focusing on non-singlet contributions involving a massive boson ($Z$ or $W$). 
The MIs relevant for diagrams featuring a single massive boson were previously determined in ref.~\cite{Pati:2025ivg}. 
However, diagrams featuring the triple vector boson vertex, many of which are subsets of those appearing in three-loop 
mixed QCD-EW corrections to Higgs production \cite{Bonetti:2017ovy}, required the evaluation of previously missing sub-topologies.
We employed state-of-the-art techniques, namely the IBP reduction to map scalar integrals to a basis of MIs, 
which were then solved using the method of differential equations. To manage the high computational cost of the IBP reduction, 
we strategically reorganized the propagator structures. By expanding the original 25 integral families from ref.~\cite{Pati:2025ivg} 
into a set of 61 families and implementing a localized reduction strategy, we significantly optimized the reduction process.
All MIs were numerically verified with \textsc{AMFlow} results at multiple kinematic points. 
Furthermore, the IR structure of the resulting FFs was found to be in complete agreement with the universal IR poles predicted for on-shell amplitudes, providing a robust cross-check of our results.

Given that mixed QCD-EW corrections are critical in specific kinematic regions, extending the precision frontier 
to this order is vital for matching the accuracy of current and future collider data. These form factors constitute 
fundamental building blocks for cross-sections, encoding the virtual corrections necessary for stable theoretical predictions. 
While this work addresses the non-singlet sector, the singlet terms, arising from diagrams with two separate Dirac traces 
involving non-trivial $\gamma_5$ structures, will be presented in a future study. 
Together, these developments enhance the theoretical framework for the DY process and 
enhance the discovery potential of high-energy collider experiments.

\section*{Acknowledgements}
We would like to thank S. Moch, A. Saha and A. Vicini for fruitful discussions. 
N.R. is partially supported by the SERB-SRG under Grant No. SRG/2023/000591.

\bibliography{main} 
\bibliographystyle{JHEP}

\end{document}